\documentclass[preprintnumbers,showpacs,amsmath,amssymb,floatfix,prd,twocolumn,superscriptaddress,nofootinbib]{revtex4}
\usepackage{bbm}
\usepackage{graphicx}
\usepackage{epsfig}
\usepackage{bm}
\usepackage{amsfonts}
\usepackage{epstopdf}

\begin{document}

\title{Quantum Spring from the Casimir Effect}

\author{Chao-Jun Feng}
\email{fengcj@shnu.edu.cn} \affiliation{Shanghai United Center for Astrophysics (SUCA), \\ Shanghai Normal University,
    100 Guilin Road, Shanghai 200234, P.R.China}

\author{Xin-Zhou Li}
\email{kychz@shnu.edu.cn} \affiliation{Shanghai United Center for Astrophysics (SUCA),  \\ Shanghai Normal University,
    100 Guilin Road, Shanghai 200234, P.R.China}

\begin{abstract}
The Casimir effect arises not only in the presence of material boundaries but also in space with nontrivial topology.
In this paper, we choose a topology of the flat $(D+1)$-dimensional spacetime, which causes the helix boundary
condition for a Hermitian massless scalar field. Especially, Casimir effect for a massless scalar field on the helix
boundary condition is investigated in two and three dimensions by using the zeta function techniques. The Casimir force
parallel to the axis of the helix behaves very much like the force on a spring that obeys the Hooke's law when the
ratio $r$ of the pitch to the circumference of the helix is small, but in this case, the force comes from a quantum
effect, so we would like to call it \textit{quantum spring}. When $r$ is large, this force behaves like the Newton's
law of universal gravitation in the leading order. On the other hand, the force perpendicular to the axis decreases
monotonously with the increasing of the ratio $r$. Both forces are attractive and their behaviors are the same in two
and three dimensions.

\end{abstract}


 \maketitle


\section{Introduction}\label{sec:intro}
Since the first work on Casimir effect performed by Casimir \cite{Casimir:1948dh}, it has been extensively studied
\cite{Plunien:1986ca} for more than 60 years. Essentially, the casimir effect is a polarization of the vacuum of some
quantized fields, and it may be thought of as the energy due to the distortion of the vacuum. Such a distortion may be
caused either by the presence of boundaries in the space-time manifold or by some background field like the gravity.
Early works on the gravity effect were performed by Utiyama and DeWitt, see
ref.\cite{Utiyama:1962sn}\cite{DeWitt:1975ys}. In history, Casimir firstly predicts the effect of the boundaries and he
found that there is an attractive force acting on two conducting plan-parallel plates in vacuum. Since the last decade,
the Casimir effect has been paid more attention due to the development of precise measurements \cite{Decca:2007yb}, and
it has been applied to the fabrication of microelectromechanical systems (MEMS)\cite{MEMS}. Recently, some new methods
have developed for computing the Casimir energy between a finite number of compact objects \cite{Emig:2007cf}.

The nature of the Casimir force may depend on (i) the background field, (ii) the spacetime dimensionality, (iii) the
type of boundary conditions, (iv) the topology of spacetime, (v) the finite temperature. The most evident example of
the dependence on the geometry is given by the Casimir effect inside a rectangular box \cite{Plunien:1986ca, Lukosz}.
The detailed calculation of the Casimir force inside a D-dimensional rectangular cavity was shown in \cite{Li}, in
which the sign of the Casimir energy depends on the length of the sides. The Casimir force arises not only in the
presence of material boundaries, but also in spaces with nontrivial topology. For example, we get the scalar field on a
flat manifold with topology of a circle $S^1$. The topology of $S^1$ causes the periodicity condition
$\phi(t,0)=\phi(t,C)$, where $C$ is the circumference of $S^1$, imposed on the wave function which is of the same kind
as those due to boundary. Similarly, the antiperiodic conditions can be drawn on a M\"obius strip. The $\zeta$-function
regularization procedure is a very powerful and elegant technique for the Casimir effect. Rigorous extension of the
proof of Epstein $\zeta$-function regularization has been discussed in \cite{Elizalde}. Vacuum polarization in the
background of on string was first considered in \cite{Helliwell:1986hs}. The generalized $\zeta$-function has many
interesting applications, e.g., in the piecewise string \cite{Li:1990bz}. Similar analysis has been applied to
monopoles \cite{BezerradeMello:1999ge}, p-branes \cite{Shi:1991qc} or pistons \cite{Zhai}.

As we have known, there are many things that look like the spring, for instance, DNA has the helix structure in our
cells. Thus, it is interesting to find the effect of the helix configuration presenting in the space-time manifold for
quantum fields and as far as we know, no one has considered this configuration before. In this paper, we have
investigated the Casimir effect for a massless scalar field on the circular helix structure in two (2D) and three (3D)
dimensions by using the zeta function techniques, which is a very useful and elegant technique in regularizing the
vacuum energy.

In next section we have calculated the Casimir energy and force by imposing the helix boundary conditions and we find
that the behavior of the force parallel to the axis of the helix is very much like the force on a spring that obeys the
Hooke's law in mechanics when the $r\ll1$, which is the ratio of the pitch $h$ to the circumference $a$ of the helix.
However, in this case, the force comes from a quantum effect, and so we would like to call the helix structure as a
\textit{quantum spring}. When $r$ is large, this force behaves like the Newton's law of universal gravitation in the
leading order and vanishes when $r$ goes to the infinity. The magnitude of this force has a maximum values at $r = 0.5$
(2D) or near $r \approx 0.494$ (3D). On the other hand, the force perpendicular to the axis decreases monotonously with
the increasing of the ratio $r$. Both forces are attractive and their behaviors are the same in two and three
dimensions. We will give some discussions and conclusions in the last section.

\section{Evaluation of the Casimir energy}\label{sec:casimir}

\subsection{Topology of the flat (D+1)-dimensional spacetime}

As mentioned in Section \ref{sec:intro}, the Casimir effect arise not only in the presence of material boundaries, but
also in spaces with nontrivial topology. For example, we get the scalar field on a flat manifold with topology of a
circle $S^1$. The topology of $S^1$ causes the periodicity condition $\phi(t,0)=\phi(t,C)$. Before we consider
complicated cases in the flat spacetime, we have to discuss the lattices.

A lattice $\Lambda$ is defined as a set of points in a flat (D+1)-dimensional spacetime $\mathcal{M}^{D+1}$, of the
form
\begin{equation}\label{lattice}
    \Lambda = \left\{ ~ \sum_{i=0}^{D} n_i \mathbbm{e}_i ~|~ n_i \in \mathcal{Z} ~\right\} \,,
\end{equation}
where $\{\mathbbm{e}_i\}$ is a set of basis vectors of $\mathcal{M}^{D+1}$. In terms of the components $v^i$ of vectors
$\mathbb{V} \in \mathcal{M}^{D+1} $, we define the inner products as
\begin{equation}\label{inner prod}
    \mathbb{V} \cdot \mathbb{W} = \epsilon(a)v^iw^j\delta_{ij} \,,
\end{equation}
with $\epsilon(a)=1$ for $i=0$, $\epsilon(a)=-1$ for otherwise. In the $x^1-x^2$ plane, the sublattice
$\Lambda''\subset\Lambda'\subset\Lambda$ are
\begin{equation}\label{sub1}
   \Lambda' = \left\{ ~  n_1 \mathbbm{e}_1 + n_2 \mathbbm{e}_2 ~|~ n_{1,2} \in \mathcal{Z} ~\right\} \,,
\end{equation}
and
\begin{equation}\label{sub2}
   \Lambda'' = \left\{ ~  n(\mathbbm{e}_1 + \mathbbm{e}_2) ~|~ n \in \mathcal{Z} ~\right\} \,.
\end{equation}

The unit cylinder-cell is the set of points
\begin{eqnarray}
 \nonumber
   U_c &=& \bigg\{\mathbb{X} = \sum_{i=0}^{D}x^i \mathbbm{e}_i ~|~ 0\leq x^1 < a,
 -h\leq x^2 < 0 , \\ && -\infty <x^0<\infty, -\frac{L}{2} \leq x^T\leq \frac{L}{2}\bigg\} \,,\label{cell}
\end{eqnarray}
where $T = 3,\cdots, D$. When $L\rightarrow\infty$, it contains precisely one lattice point (i.e. $\mathbb{X} = 0$),
and any vector $\mathbb{V}$ has precisely one "image" in the unit cylinder-cell, obtained by adding a sublattice vector
to it.

\begin{figure}[h]
\begin{center}
\includegraphics[width=0.4\textwidth]{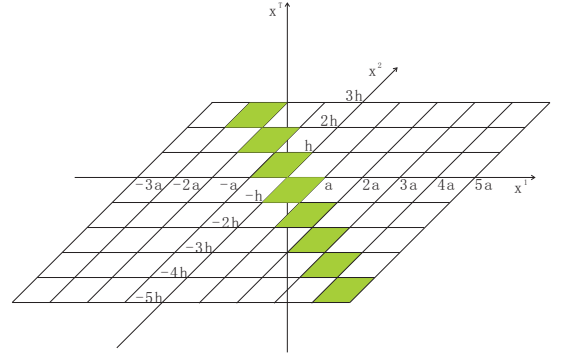}
\caption{\label{fig::lattice} The helix boundary condition can be induced by the topology of spacetime.}
\end{center}
\end{figure}

In this paper, we choose a topology of the flat (D+1)-dimensional sapcetime: $U_c\equiv U_c +\mathbbm{u}, \mathbbm{u}
\in \Lambda''$, see Fig.\ref{fig::lattice}. This topology causes the helix boundary condition for a Hermitian massless
scalar field
\begin{equation}\label{helxi boundary condition}
   \phi(t, x^1 + a, x^2, x^T) =  \phi(t, x^1 , x^2+h, x^T) \,,
\end{equation}
where, if $a=0$ or $h=0$, it returns to the periodicity boundary condition.

In calculations on the Casimir effect, extensive use is made of eigenfunctions and eigenvalues of the corresponding
field equation. A Hermitian massless scalar field $\phi(t, x^\alpha, x^T)$ defined in a  (D+1)-dimensional flat
spacetime satisfies the free Klein-Gordon equation:
\begin{equation}\label{eom}
    \left(\partial_t^2 - \partial_i^2\right)\phi(t, x^\alpha, x^T) = 0 \,,
\end{equation}
where $i=1,\cdots, D; \alpha=1,2; T=3,\cdots, D$. Under the boundary condition (\ref{helxi boundary condition}), the
modes of the field are then
\begin{equation}\label{modes}
    \phi_{n}(t, x^\alpha, x^T)= \mathcal{N} e^{-i\omega_nt+ik_x x+ik_z z + ik_Tx^T }\,,
\end{equation}
where $\mathcal{N}$ is a normalization factor and $x^1=x, x^2=z$, and we have
\begin{equation}\label{energy}
    w_n^2 = k_{T}^2 + k_x^2 + \left( -\frac{2\pi n}{h}+\frac{k_x}{h}a \right)^2 = k_{T}^2 + k_z^2 + \left( \frac{2\pi n}{a}+\frac{k_z}{a}h
    \right)^2 \,.
\end{equation}
Here, $k_x$ and $k_z$ satisfy
\begin{equation}\label{kxkz}
    a k_x - hk_z = 2n\pi\,, (n=0,\pm1,\pm2,\cdots) \,.
\end{equation}
In the ground state (vacuum), each of these modes contributes an energy of $w_n/2$. The energy density of the field is
thus given by
\begin{eqnarray}
\nonumber
  &E^{D+1}& = \frac{1}{2 a}
  \int \frac{d^{D-1}k}{(2\pi)^{D-1}} \sum_{n=-\infty}^{\infty} \sqrt{k_T^2 + k_z^2 + \left( \frac{2\pi n}{a}+\frac{k_z}{a}h
    \right)^2  } \,, \\&&\label{tot energy}
\end{eqnarray}
where we have assumed $a\neq 0$ without losing generalities.

\subsection{Massless scalar field in $2+1$ dimension}
\begin{figure}[h]
\begin{center}
\includegraphics[width=0.4\textwidth]{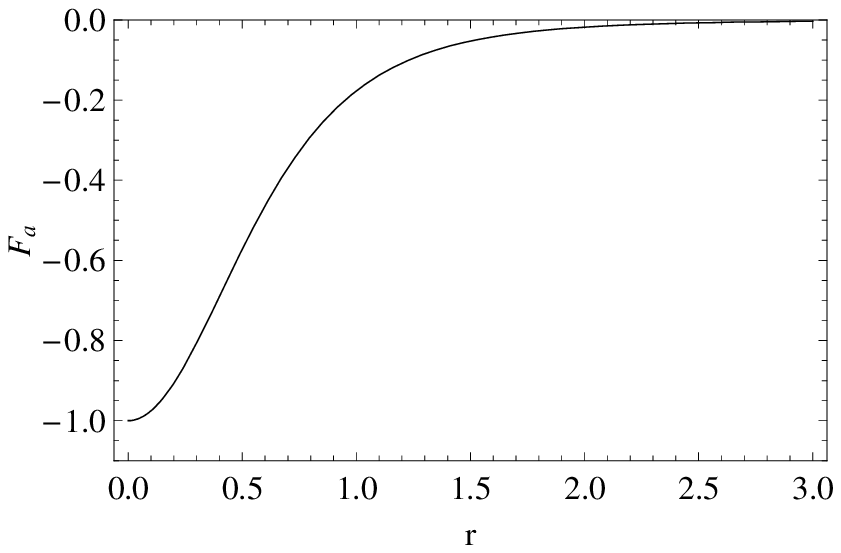}
\qquad
\includegraphics[width=0.4\textwidth]{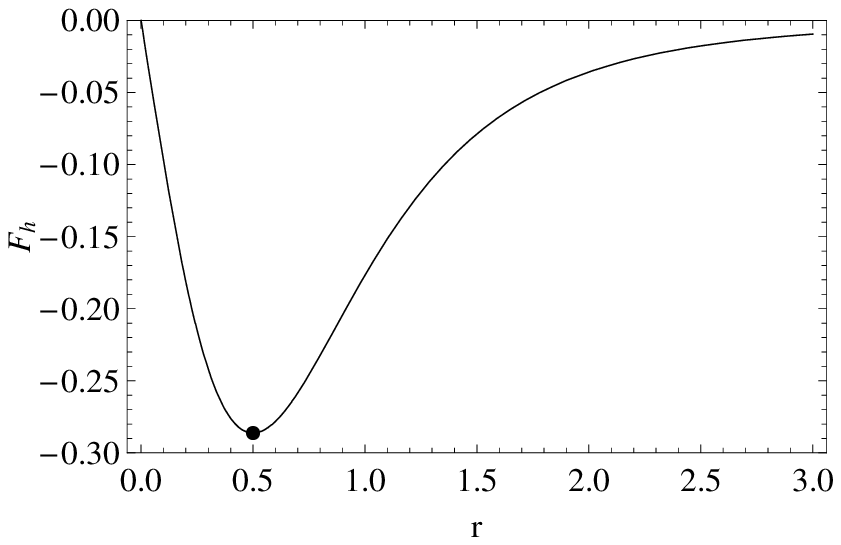}
\caption{\label{fig::force2d}The Casimir force on the $x$ (left) and $z$ (right) direction in the unit $3\zeta(3)/(2\pi
a^4)$ \textit{vs.} the ratio $r$ in $2+1$ dimension. The point corresponds to the maximum magnitude of the force at
$r=0.5$.}
\end{center}
\end{figure}

In the $2+1$ dimensional spacetime, we have the following boundary condition to mimic the helix structure:
\begin{equation}\label{bdc}
   \phi(t, x+a, z)= \phi(t,x,z+h)\,,
\end{equation}
where $h$ is regarded as the pitch of the helix, and we call this condition the helix boundary condition. One can see
from eq.(\ref{bdc}) that it would return to the cylindrical boundary conditions when $h$ vanishes and for $h\neq 0$,
the whole system(the spring) does not have the cylindrical symmetry. Therefore, the vacuum energy  density is given by
\begin{equation}
E(a,h) = \frac{1}{2a}\int_{-\infty}^{\infty} \frac{dk}{2\pi} \sum_{n=-\infty}^{\infty}\sqrt{k^2 + \bigg(\frac{2\pi
n}{a}+\frac{k}{a} h\bigg)^2} \,,
\end{equation}
which is divergent, so we should regularize it to get a finite result. There many regularization method could be used
to deal with the divergence, but in this paper we would like to use the zeta function techniques, which is a very
useful and elegant technique in regularizing the vacuum energy. To use the $\zeta$-function regularization, we define
$\mathcal{E}(s)$ as
\begin{equation}\label{es}
    \mathcal{E}(a,h;s) = \frac{\sqrt{\gamma}}{\pi a}\sum_{n=1}^{\infty}\int_{0}^{\infty} dk\left(k^2 + 1\right)^{-s/2}\left(\frac{2\pi
    n}{a\gamma}\right)^{1-s} \,,
\end{equation}
for $Re(s)>1$ to make a finite result provided by the $k$ integration, and here we have defined
\begin{equation}\label{gamma}
     \gamma \equiv 1+ r^2 \,, \quad  r = \frac{h}{a}\,.
\end{equation}
We will see in the following that the analytic continuation to the complex $s$ plane is well defined at $s=-1$. Thus,
the regularized Casimir energy density is $E_R(a,h)=\mathcal{E}(a,h;-1)$. After integrating $k$ in eq.(\ref{es}), see
Appendix \ref{int},  we get
\begin{equation}\label{es2}
   \mathcal{E}(a,h;s) =  \frac{1}{2 a}\sqrt{\frac{\gamma}{\pi}}\left(\frac{2\pi }{a\gamma}\right)^{1-s}
    \frac{ \Gamma\left(\frac{s-1}{2}\right)}{ \Gamma\left(\frac{s}{2}\right)}
    \zeta(s-1) \,,
\end{equation}
where $\zeta(s)$ is the Riemann zeta function. The value of the analytically continued zeta function can be obtained
from the reflection relation
\begin{equation}\label{rel}
    \Gamma\left(\frac{s}{2}\right)\zeta(s) = \pi^{s-\frac{1}{2}} \Gamma\left(\frac{1-s}{2}\right)\zeta(1-s)\,.
\end{equation}
Taking $s=-1$, we get
\begin{equation}
   \lim_{s\rightarrow-1} \Gamma\left(\frac{s-1}{2}\right)\zeta(s-1) =  \frac{\zeta(3)}{2\pi^2}\,,
\end{equation}
then we have
\begin{equation}\label{r e}
    E_R(a,h)  = -\frac{\zeta(3)}{2\pi a^3}\gamma^{-3/2} =-\frac{\zeta(3)}{2\pi a^3}\bigg(1+r^2\bigg)^{-3/2} \,,
\end{equation}
where we have used $\Gamma(-1/2) = -2\sqrt{\pi}$ and if $r=0$, it come back to the cylindrical case  with periodical
boundary, see eq.(\ref{bdc}). The Casimir force on the $x$ direction of the helix  is
\begin{equation}\label{force1 a}
    F_a = -\frac{\partial E_R(a,h)}{\partial a} = - \frac{3\zeta(3)}{2\pi a^4} \bigg(1+r^2\bigg)^{-5/2},
\end{equation}
which is always an attractive force and the magnitude of the force monotonously decreases with the increasing of the
ratio $r$. Once $r$ becomes large enough, the force can be neglected. While, the Casimir force on the $z$ direction is
\begin{equation}\label{force1 h}
    F_h = -\frac{\partial E_R(a,h)}{\partial h} = - \frac{3\zeta(3)}{2\pi a^4} \frac{r}{(1+r^2)^{5/2}},
    \,.
\end{equation}
which has a maximum magnitude at $r =0.5$. When $r<0.5$, the magnitude of the force increases with the increasing of
$r$ until $r=0.5$, and the force is almost linearly depending on $r$ when $r\ll1$. So, it is just like the force on a
spring complying with the Hooke's law, but in this case, the force  originates from the quantum effect, namely, the
Casimir effect. Once $r>0.5$, the magnitude of the force decreases with the increasing of $r$.  To illustrate the
behavior of the Casimir force in this case, we plot them for each direction in Fig.\ref{fig::force2d}.

It should be noticed that in  Fig.\ref{fig::force2d}, the behavior of the forces are different with respect to the
ratio $r$, but this dose not conflict with the relation (\ref{energy}), which shows that labeling the axes is a matter
of convention, namely the final result should have the the symmetry of $a\leftrightarrow h$. The reason is the
following, eq.(\ref{r e}) could be rewritten in terms of $a$ and $h$:
\begin{equation}
    E_R(a,h)  = -\frac{\zeta(3)}{2\pi}\bigg(a^2+h^2\bigg)^{-3/2} \,,
\end{equation}
which respects the symmetry of $a\leftrightarrow h$ in deed. And, one can easily see that eqs. (\ref{force1 a}) and
(\ref{force1 h}) are also under this symmetry, if one rewritten these equations as
\begin{eqnarray}
  F_a &=& - \frac{3\zeta(3)}{2\pi} \frac{a}{(a^2+h^2)^{5/2}}\,, \\
  F_h &=&  - \frac{3\zeta(3)}{2\pi} \frac{h}{(a^2+h^2)^{5/2}} \,,
\end{eqnarray}
which are all consistent with the relation (\ref{energy}).

\subsection{Massless scalar field in $3+1$ dimension}

\begin{figure}[h]
\begin{center}
\includegraphics[width=0.3\textwidth]{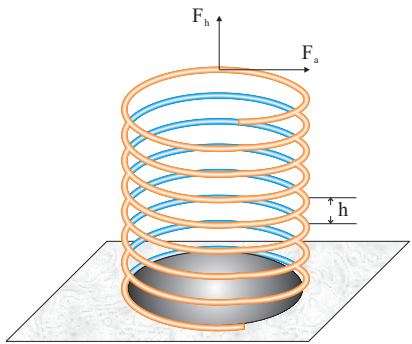}
\caption{\label{fig::spring} Illustration of the \textit{Quantum spring}.}
\end{center}
\end{figure}

\begin{figure}[h]
\begin{center}
\includegraphics[width=0.4\textwidth]{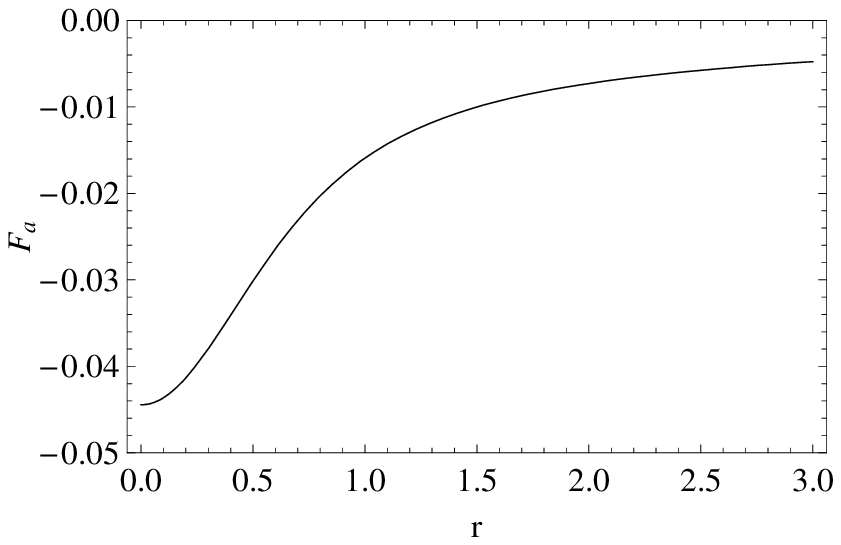}
\qquad
\includegraphics[width=0.4\textwidth]{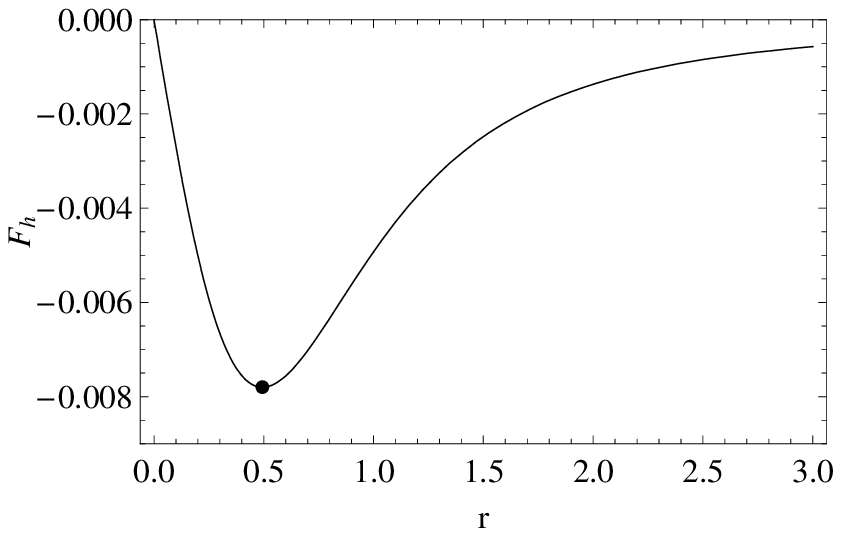}
\caption{\label{fig::force3d}The Casimir force on the $x$ (left) and $z$ (right) direction in the unit $\pi^2/a^5$
\textit{vs.} the ratio $r$ in $3+1$ dimension. The point corresponds to the maximum magnitude of the force at
$r=r_0\approx0.494$.}
\end{center}
\end{figure}

As in the $2+1$ dimension case, the vacuum energy  density in $3+1$ dimention is given by
\begin{equation}
E(a,h) = \frac{1}{2a}\int_{-\infty}^{\infty} \frac{dk_ydk_z}{(2\pi)^2} \sum_{n=-\infty}^{\infty}\sqrt{k^2 +
\bigg(\frac{2\pi n}{a}+\frac{k_z}{a} h\bigg)^2} \,,
\end{equation}
where $k^2 = k^2_y + k^2_z$. Again, to use the $\zeta$-function regularization, we define $\mathcal{E}(s)$ as
\begin{eqnarray}
\nonumber
  \mathcal{E}(a,h;s) &=& \frac{1}{4\pi^2 a}\sum_{n=1}^{\infty}\int_0^{2\pi}d\theta \sqrt{\tilde\gamma} \\
  &&\cdot \int_{0}^{\infty} kdk\left(k^2 + 1\right)^{-s/2}\left(\frac{2\pi n}{a\tilde\gamma}\right)^{2-s} \,,\label{es 31}
\end{eqnarray}
and  for $Re(s)>1$, and we have defined
\begin{equation}\label{gamma 31}
    \tilde\gamma \equiv 1+ r^2\cos^2\theta\,.
\end{equation}
where $\cos\theta = k_z/k$ and $r$ is still the ratio of $h$ to $a$ defined in eq.(\ref{gamma}). We will see in the
following that the analytic continuation to the complex $s$ plane is also well defined at $s=-1$ in this case. Thus,
the regularized Casimir energy density is $E_R(a,h)=\mathcal{E}(a,h;-1)$. After integrating $k$ and $\theta$ in
eq.(\ref{es 31}),  see Appendix \ref{inte}, we get
\begin{equation}\label{es2 31}
   \mathcal{E}(a,h;s) = - \frac{\zeta(s-2)}{2\pi(2-s) a}\left(\frac{2\pi}{a}\right)^{2-s}~{}_2F_1\bigg(\frac{3}{2}-s,\frac{1}{2},1;-r^2\bigg)\,,
\end{equation}
Taking $s=-1$, we get $\zeta(-3)=\frac{1}{120}$ from (\ref{rel}), and then
\begin{equation}\label{r e 31}
    E_R(a,h)  = -\frac{\pi^2}{90 a^4} ~{}_2F_1\bigg(\frac{5}{2},\frac{1}{2},1;-r^2\bigg)\,.
\end{equation}
Therefore, the Casimir force on the $x$ direction of the helix  is
\begin{eqnarray}
\nonumber
  F_a &=& -\frac{\partial E_R(a,h)}{\partial a} \\
\nonumber
  &=&-\frac{2\pi^2}{45 a^5}
   \bigg[~{}_2F_1\bigg(\frac{5}{2},\frac{1}{2},1;-r^2\bigg) \\
\nonumber
   & &- \frac{5r^2}{8}~{}_2F_1\bigg(\frac{7}{2},\frac{3}{2},2;-r^2\bigg)\bigg] \,,\\&&\label{force a}
\end{eqnarray}
which is always attractive and its magnitude  monotonously decreases with the increasing of the ratio $r$. By the
definition of the hypergeometric function, we can expand eq.(\ref{force a}) up to arbitrary orders of $r$, thus for
small $r$, we get
\begin{equation}
    F_a|_{r\ll1} = -\frac{2\pi^2}{45 a^5}\left[1-\frac{15}{8}r^2 +\mathcal{O}(r^4)\right] \,,
\end{equation}
while for large $r$, we asymptotically expand eq.(\ref{force a}) as
\begin{equation}
    F_a|_{r\gg1} =  -\frac{2\pi^2}{45 a^5}\left[\frac{1}{\pi r} + \frac{1}{12\pi r^3} +
    \mathcal{O}\left(\frac{1}{r^5}\right)\right]\,,
\end{equation}
up to $\mathcal{O}(r^{-5})$. Then, it is clear to see that, the force will be vanished when $r$ goes to infinity. On
the other hand, the Casimir force on the $z$ direction is
\begin{equation}\label{force h}
    F_h = -\frac{\partial E_R(a,h)}{\partial h} = -\frac{\pi^2r}{36 a^5}
    ~{}_2F_1\bigg(\frac{7}{2},\frac{3}{2},2;-r^2\bigg)\,,
\end{equation}
and for $r\ll1$ and $r\gg1$, we respectively have
\begin{equation}
    F_h|_{r\ll1} = -\frac{\pi^2}{36 a^5}\left[r-\frac{21}{8}r^3 +\mathcal{O}(r^5)\right]\,,
\end{equation}
and
\begin{equation}
    F_h|_{r\gg1} = -\frac{\pi^2}{36 a^5}\left[\frac{8}{15\pi r^2} + \frac{2}{5\pi r^4} +
    \mathcal{O}\left(\frac{1}{r^6}\right)\right] \,.
\end{equation}
Therefore, for small $r$, the force  linearly depends on $r$, namely,
\begin{equation}
    F_h = - K r \,, \quad K = \frac{\pi^2}{36 a^5}\,,  \quad (r\ll1) \,,
\end{equation}
which is very much like a spring obeying the Hooke's law with spring constant $K$ in classical mechanics, but in this
case, the force comes from a quantum effect, and we would like to call it \textit{quantum spring}, see
Fig.\ref{fig::spring}. When $r$ is large, the force behaves like the Newton's law of universal gravitation, i.e.
$F_h\sim -1/r^2$ in the leading order. Furthermore, there exists a maximum magnitude of the force $|F_h|_{max} $ when
$r$ takes a critical value $r_0 \approx 0.494$, which satisfy the following equation
\begin{equation}
    4~{}_2F_1\left(7/2,3/2,2;-r_0^2\right) - 21~
    {}_2F_1\left(9/2,5/2,3;-r_0^2\right)r_0^2 =0 \,.
\end{equation}
To illustrate the behavior of the forces on the helix, we plot them for each direction in Fig.\ref{fig::force3d}.

\section{Conclusion and discussion}

In conclusion, we have investigated the Casimir effect with a helix configuration in two and three dimensions, and it
can be easily generalized to high dimensions. We find that the force parallel to the axis of the helix has a particular
behaviors that the Casimir force in the usual case do no possesses. It behaves very much like the force on a spring
that obeys the Hooke's law in mechanics when  $r\ll1$, and like the Newton's law of universal gravitation when $r\gg1$.
Furthermore, the The magnitude of this force has a maximum values at $r = 0.5$ (2D) or near $r \approx 0.494$ (3D). So,
we would like to call this helix configuration as a \textit{quantum spring}, see Fig.\ref{fig::spring}. On the other
hand, the force perpendicular to the axis decreases monotonously with the increasing of the ratio $r$. Both forces are
attractive and their behaviors are the same in two and three dimensions.

\begin{figure}[h]
\begin{center}
\includegraphics[width=0.5\textwidth]{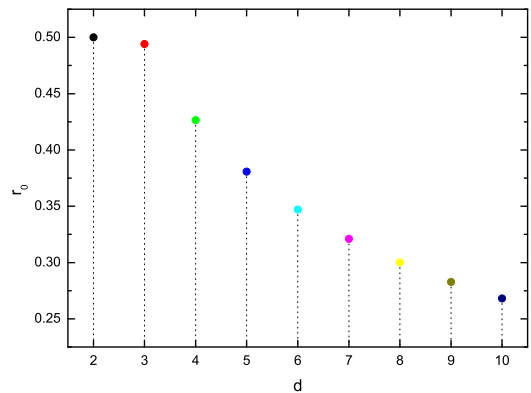}
\caption{\label{fig::rd}The critical value $r_0$ \textit{vs}. the space dimension $d$.}
\end{center}
\end{figure}

It should be noticed that, the critical value $r_0$, at witch the magnitude of the force gets its maximum value depends
on the space-time dimensions. On a general $D+1$-dimensional ($D\ge3$) flat space-time manifold, the Casimir energy
density on the helix is roughly given by
\begin{equation}
    E(a,h)\sim -  ~{}_2F_1\bigg(d-\frac{1}{2},\frac{1}{2},1;-r^2\bigg)a^{-(d+1)} \,,
\end{equation}
up to some coefficient. Then, the Casimir force on the $z$ direction is roughly
\begin{equation}
    F_h \sim -  r ~{}_2F_1\bigg(d+\frac{1}{2},\frac{3}{2},2;-r^2\bigg)a^{-(d+2)} \,,
\end{equation}
thus the critical value $r_0$ satisfies
\begin{eqnarray}
\nonumber
    && 4~{}_2F_1\left(d+1/2,3/2,2;-r_0^2\right) \\
    &&- 3(2d+1)~
    {}_2F_1\left(d+3/2,5/2,3;-r_0^2\right)r_0^2 =0 \,,
\end{eqnarray}
which can not be exactly solved but one can numerically calculate the critical point $r_0$. In Fig.{\ref{fig::rd}}, we
have shown the dependence of $r_0$ on the space dimension $d$ from two to ten dimensions.

In this paper, we have considered the massless scalar field, and one can easily generalize it to a massive scalar
field. As is known that the Casimir effect disappears as the mass of the field goes to infinity since there are no more
quantum fluctuation in this limit, but of course, how the Casimir force varies as the mass changes is worth studying
\cite{Bordag:2001qi}, and we will study it in our further work \cite{fur}, in which we will also consider the Casimir
effect of the electromagnetic field in the helix configuration. Since this \textit{quantum spring} effect may be
detected in the laboratory and be applied to the microelectromechanical system, we suggest to do the experiment to
verify our results. It should be noticed that, in the experiment or the real application, the spring  like
Fig.\ref{fig::spring} should be soft, which means the force coming from the classical mechanics could be small enough,
and the quantum effect dominates the behavior of the spring.

\acknowledgments

This work is supported by National Education Foundation of China grant No. 2009312711004 and Shanghai Natural Science
Foundation, China grant No. 10ZR1422000.

\appendix
\section{The $k$ integration in $2+1$ dimension}

The integration  in eq.(\ref{es}) is given by
\begin{eqnarray}
\nonumber
  \mathcal{I}(s) &=& \int_{0}^{\infty} dk\left(k^2 + 1\right)^{-s/2}
    =\int_{0}^{\infty} {}_2F_1\left(\frac{s}{2},b,b;-k^2\right)dk \\
    &=& \frac{1}{2}\int_{0}^{\infty}
    {}_2F_1\left(\frac{s}{2},b,b;-z\right)z^{-1/2}dz \,,\label{int}
\end{eqnarray}
where we have used $(1+z)^\alpha = {}_2F_1\left(-\alpha,b,b;-z\right)$, and ${}_pF_q$ is  hypergeometric functions. By
using \cite{integ}
\begin{eqnarray}
\nonumber
  \int^\infty_0 {}_2F_1(a,b,c;-z) z^{-t-1}dz
  &=&\frac{\Gamma(a+t)\Gamma(b+t)\Gamma(c)\Gamma(-t)}{\Gamma(a)\Gamma(b)\Gamma(c+t)} \\
  &&
\end{eqnarray}
where $\Gamma(a)$ is  Gamma functions, we get
\begin{equation}\label{int3}
        \mathcal{I}(s)= \frac{ \sqrt{\pi }}{2} \frac{ \Gamma\left(\frac{s-1}{2}
    \right)}{ \Gamma\left(\frac{s}{2}\right)}\,,
\end{equation}
where we have used $\Gamma(1/2)=\sqrt{\pi}$.

\section{The $k$ and $\theta$ integration in $3+1$ dimension}\label{inte}

The integration  in eq.(\ref{es 31}) is given by
\begin{eqnarray}
 \nonumber
   \mathcal{I}(s) &=& \int_0^{2\pi}d\theta ~\tilde\gamma^{s-3/2} \int_{0}^{\infty} kdk\left(k^2 + 1\right)^{-s/2}\\
   \nonumber
     &=& \frac{1}{2-s}\int_0^{2\pi}d\theta ~\tilde\gamma^{s-3/2}(k^2+1)^{1-\frac{s}{2}}\bigg|_{k=0}^\infty \\
   \nonumber
     &=& -\frac{4}{2-s}\int_0^{\pi/2}d\theta ~\bigg(1+r^2\cos^2\theta\bigg)^{s-3/2} \\
   \nonumber
     &=& -\frac{2}{2-s}\int_0^{1}x^{-1/2}(1-x)^{-1/2} \\
   \nonumber && \cdot~{}_2F_1\bigg(\frac{3}{2}-s,\frac{1}{2},\frac{1}{2};-r^2 x\bigg) dx\\
     &=& -\frac{2\pi}{2-s} ~{}_2F_1\bigg(\frac{3}{2}-s,\frac{1}{2},1;-r^2\bigg)\,,
\end{eqnarray}
where we have defined $x=\cos^2\theta$ and we have used \cite{integ}
\begin{eqnarray}
\nonumber
    &&\int_0^1 (1-x)^{\mu-1}x^{\nu-1} {}_pF_q(a_1,\cdots,a_p; \nu, b_2, \cdots,b_q; ax)dx \\
    &=& \frac{\Gamma(\mu)\Gamma(\nu)}{\Gamma(\mu+\nu)}{}_pF_q(a_1,\cdots,a_p; \mu+\nu, b_2, \cdots,b_q; a) \,.
\end{eqnarray}


\begin{thebibliography}{999}

\bibitem{Casimir:1948dh}
  H.~B.~G.~Casimir,
  Indag.\ Math.\  {\bf 10}, 261 (1948)
  [Kon.\ Ned.\ Akad.\ Wetensch.\ Proc.\  {\bf 51}, 793 (1948\ FRPHA,65,342-344.1987\ KNAWA,100N3-4,61-63.1997)].

\bibitem{Plunien:1986ca}
  M.~Bordag, G.~L.~Klimchitskaya, U.~Mohideen and V.~M.~ Mostepanenko, \textit{Advances in the Casimir Effect}, Oxford
  University Press, 2009.


\bibitem{Utiyama:1962sn}
  R.~Utiyama and B.~S.~DeWitt,
  J.\ Math.\ Phys.\  {\bf 3}, 608 (1962).

\bibitem{DeWitt:1975ys}
  B.~S.~DeWitt,
  Phys.\ Rept.\  {\bf 19}, 295 (1975).

\bibitem{Decca:2007yb}
  R.~S.~Decca, D.~Lopez, E.~Fischbach, G.~L.~Klimchitskaya, D.~E.~Krause and V.~M.~Mostepanenko,
  Phys.\ Rev.\  D {\bf 75}, 077101 (2007)
  [arXiv:hep-ph/0703290].

\bibitem{MEMS}
  F.~M.~Serry, D.~Walliser, and G.~J.~Maclay, J.Microelectromech.Syst. {\bf4}, 193 (1995), \\
  H.~B.~Chan, V.~A.~Aksyuk, R.~N.~Kleiman, D.~J.~Bishop, and F.~Capasso, Science {\bf291}, 1941 (2001).

\bibitem{Emig:2007cf}
  T.~Emig, N.~Graham, R.~L.~Jaffe and M.~Kardar,
  Phys.\ Rev.\ Lett.\  {\bf 99}, 170403 (2007)
  [arXiv:0707.1862 [cond-mat.stat-mech]].

\bibitem{Lukosz}
  W.~Lukosz, Physica {\bf 56}, 109(1971).

\bibitem{Li}
  X.~Z.~Li, H.~B.~Cheng, J.~M.~Li and X.~H.~Zhai,
  Phys.\ Rev.\  D {\bf 56}, 2155 (1997);\\
  X.~Z.~Li and X.~H.~Zhai,
  J.\ Phys.\ A {\bf34}:11053-11057, 2001.
  [arXiv:hep-th/0205225].


\bibitem{Elizalde}
  E.~Elizalde, S.~D.~Odintsov, A.~Romeo, A.~A.~Bytsenko and S.~Zerbini, \textit{Zeta Regularization Techniques with
  Applications}, World Scientific, Singapore, 1993.

\bibitem{Helliwell:1986hs}
  T.~M.~Helliwell and D.~A.~Konkowski,
  Phys.\ Rev.\  D {\bf 34}, 1918 (1986).

\bibitem{Li:1990bz}
  X.~Z.~Li, X.~Shi and J.~Z.~Zhang,
  Phys.\ Rev.\  D {\bf 44}, 560 (1991);\\
  I.~H.~Brevik, H.~B.~Nielsen and S.~D.~Odintsov,
  Phys.\ Rev.\  D {\bf 53}, 3224 (1996).

\bibitem{BezerradeMello:1999ge}
  E.~R.~Bezerra de Mello, V.~B.~Bezerra and N.~R.~Khusnutdinov,
  Phys.\ Rev.\  D {\bf 60}, 063506 (1999)
  [arXiv:gr-qc/9903006].

\bibitem{Shi:1991qc}
  X.~Shi and X.~ .~Li,
  Class.\ Quant.\ Grav.\  {\bf 8}, 75 (1991).

\bibitem{Zhai}
  X.~H.~Zhai and X.~Z.~Li,
  Phys.\ Rev.\  D {\bf 76}, 047704 (2007)
  [arXiv:hep-th/0612155]; \\
  X.~H.~Zhai, Y.~Y.~Zhang and X.~Z.~Li,
  Mod.\ Phys.\ Lett.\  A {\bf 24}, 393 (2009)
  [arXiv:0808.0062 [hep-th]];\\
  R.~M.~Cavalcanti,
  Phys.\ Rev.\  D {\bf 69}, 065015 (2004)
  [arXiv:quant-ph/0310184];\\
  M.~P.~Hertzberg, R.~L.~Jaffe, M.~Kardar and A.~Scardicchio,
  Phys.\ Rev.\ Lett.\  {\bf 95}, 250402 (2005)
  [arXiv:quant-ph/0509071].

\bibitem{integ}
I.S. Gradshteyn and I.M. Ryzhik ; Alan Jeffrey, Daniel Zwillinger, editors. \textit{Table of Integrals, Series, and
Products}, seventh edition. Academic Press, 2007. ISBN 978-0-12-373637-4 .

\bibitem{Bordag:2001qi}
  M.~Bordag, U.~Mohideen and V.~M.~Mostepanenko,
  Phys.\ Rept.\  {\bf 353}, 1 (2001)
  [arXiv:quant-ph/0106045].

\bibitem{fur}
   C.~J.~Feng and X.~Z.~Li, work in progress.

\end{thebibliography}
\end{document}